\documentclass[%
reprint,
nofootinbib,
 amsmath,amssymb,
 aps,
]
{revtex4-1}
\usepackage{epsfig,epsf,rotating,wrapfig,tabularx}
\usepackage{dcolumn}
\usepackage{bm}
\usepackage{graphicx}

\begin{document}

\preprint{APS/123-QED}

\title{Predictions on the transverse momentum spectra for charged particle production at LHC-energies from a two component model}

\author{\firstname{A.~A.}~\surname{Bylinkin}}
 \email{alexandr.bylinkin@cern.ch}
\affiliation{%
Moscow Institute of Physics and Technology, MIPT, Moscow, Russia\\
National Research Nuclear University MEPhI, Moscow, Russia
}%
\author{\firstname{N.~S.}~\surname{Chernyavskaya}}
 \email{nadezda.chernyavskaya@cern.ch}
\affiliation{%
Moscow Institute of Physics and Technology, MIPT, Moscow, Russia\\
 Institute for Theoretical and Experimental
Physics, ITEP, Moscow, Russia\\
National Research Nuclear University MEPhI, Moscow, Russia
}%
\author{\firstname{A.~A.}~\surname{Rostovtsev}}
 \email{rostov@itep.ru}
\affiliation{%
 Institute for Information Transmission Problems, IITP, Moscow, Russia
}%


\begin{abstract}
Transverse momentum spectra, $d^2\sigma/(d\eta dp_T^2)$, of charged hadron production in $pp$-collisions are considered in terms of a recently introduced two component model. The shapes of the particle distributions vary as a function of c.m.s. energy in the collision and the measured pseudorapidity interval. In order to extract predictions on the double-differential cross-sections  $d^2\sigma/(d\eta dp_T^2)$ of hadron production for future LHC-measurements the different sets of available experimental data have been used in this study. 
\end{abstract}

\pacs{Valid PACS appear here}
\maketitle

\section{Introduction}
Recently a qualitative model considering two sources of hadroproduction has been introduced~\cite{OUR1}. It was suggested to parametrize charged particle spectra by a sum of an exponential (Boltzmann-like) and a power-law $p_T$ distributions:
\begin{equation}
\label{eq:exppl}
\frac{d^2\sigma}{d\eta d p_T^2} = A_e\exp {(-E_{Tkin}/T_e)} +
\frac{A}{(1+\frac{p_T^2}{T^{2}\cdot N})^N},
\end{equation}
where  $E_{Tkin} = \sqrt{p_T^2 + M^2} - M$
with M equal to the produced hadron mass. $A_e, A, T_e, T, N$ are the free parameters to be determined by fit to the data.  The detailed arguments for this particular choice are given in~\cite{OUR1}.  The exponential term in this model is associated with thermalized production of hadrons by valence quarks and a quark-gluon cloud coupled to them. While the power-law term is related to the mini-jet fragmentation of the virtual partons (pomeron in pQCD) exchanged between two colliding partonic systems. 
 
\begin{figure}[h]
\includegraphics[width =8cm]{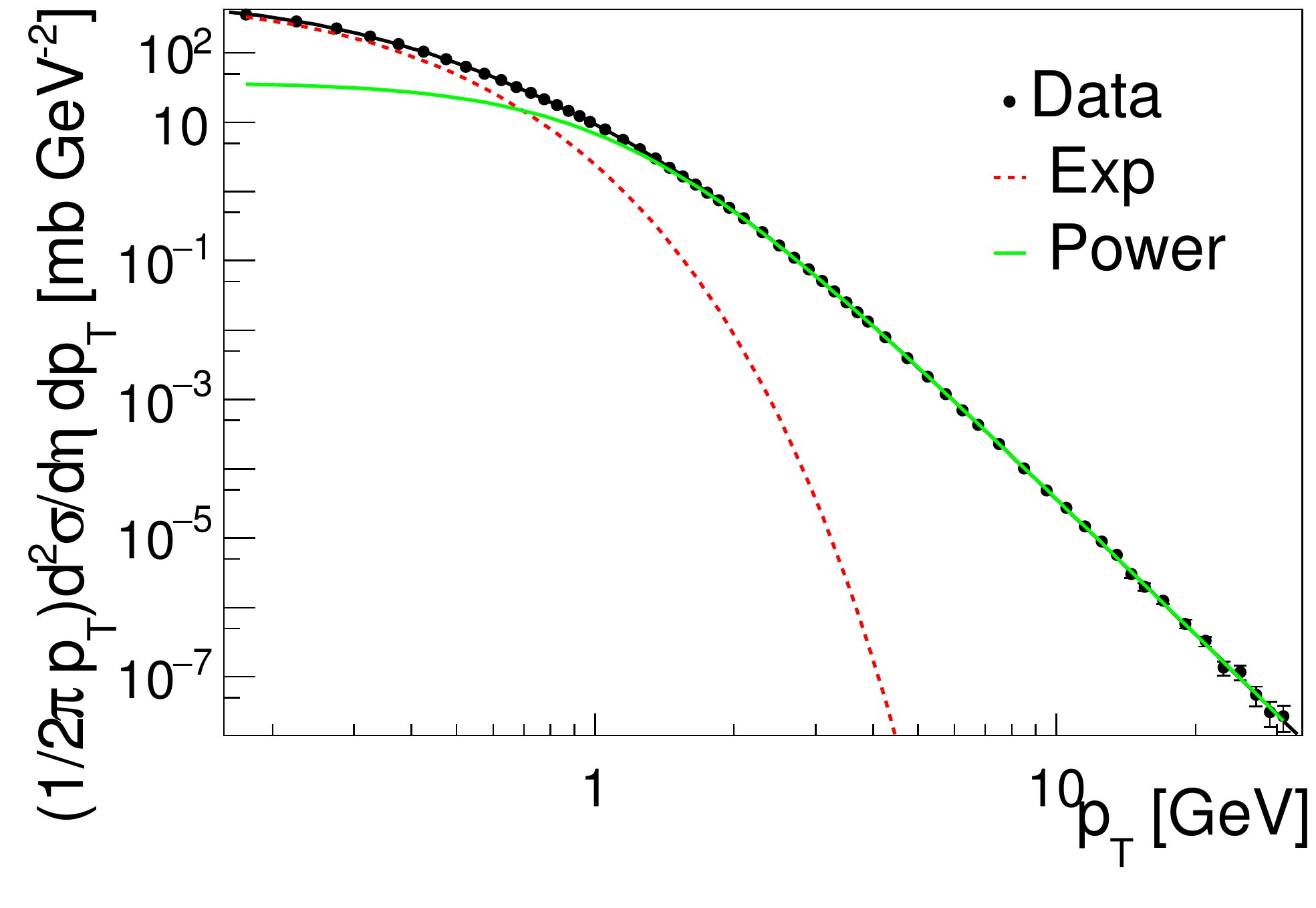}
\caption{\label{fig.0} Charge particle differential cross section $(1/2\pi p_{T}) d^2\sigma/d\eta dp_{T} $~\cite{ALICE} fitted to the function~(\ref{eq:exppl}): the red (dashed) line shows the exponential term and the green (solid) one - the power law.}
\end{figure}

A typical charged particle spectrum as a function of transverse momentum, fitted with this function~(\ref{eq:exppl}) is shown in figure~\ref{fig.0}. 
As one can see, the exponential term dominates the particle spectrum at low $p_T$ values.

This model has already been shown to predict the mean $<p_T>$ values as a function of multiplicity in a collision~\cite{OURM} and pseudorapidity distributions of charged particles~\cite{OURR}. However, the major interest of many studies in QCD is the transverse momentum spectrum itself. Therefore, in this article it is discussed how its shape varies in different experiments under various conditions. In~\cite{OUR1} it was shown that the parameters of the fit~(\ref{eq:exppl}) show a strong dependence on the collision energy. Unfortunately, due to the fact that different collaborations measure charged particle production in their own phase space and under various experimental setup, the dependences observed in~\cite{OUR1} were smeared and did not allow to make strong predictions for further measurements. Thus, an approach to correct the measurements in order to allow an accurate combination of different experimental data is proposed here.

\section{Parameter variations}

In~\cite{OURR} it was shown that two sources of hadroproduction described above, contribute to different pseudorapidity regions: while the power-law term of (\ref{eq:exppl}) prevails in the most central region ($\eta\sim0$), the exponential term dominates at high values of $\eta$. Since each collaboration presents measurements on transverse momentum spectra in various pseudorapidity intervals, these variations might explain the smearing of the dependences in~\cite{OUR1}.
The idea to study parameter variations as a function of both collision energy and pseudorapidity region has already been successfully tested in~\cite{Confin}.

\begin{figure}[h]
\includegraphics[width =8cm]{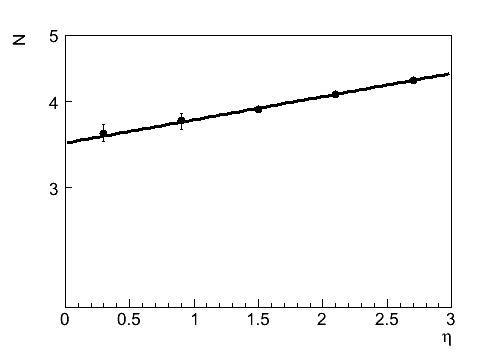}
\caption{\label{fig.1} Variation of the $N$ parameter of (\ref{eq:exppl}) obtained from the fits to the experimental data~\cite{UA1} as a function of pseudorapidity. Line shows a power-law fit of this variation.}
\end{figure}

To further study the variations of the spectra shape as a function of pseudorapidity we use the data published by the UA1 experiment~\cite{UA1} which are presented by charged particle spectra in five pseudorapidity bins, covering the total rapidity interval $|\eta|<3.0$. Figure~\ref{fig.1} shows how the parameter $N$ varies with pseudorapidity together with a power-law fit of this variation. Note, that the parameter shows a growth with pseudorapidity~\cite{OURR}, that is explained by higher thermalization of the spectra, as found in~\cite{OURR}.

This variation can be parametrized in the following way:
\begin{equation}
\label{eq.n}
N = N^0\cdot(1+ 0.06\cdot |\eta|^{1.52})
\end{equation}

where $N^0$ denotes the parameter value at $\eta\sim0$ and $\eta$ might be taken as a mean $<\eta>$ of the measured pseudorapidity interval.

Since the variations of the parameter as function of pseudorapidity have been found, it is desirable to exclude its influence when studying the dependences of $N$ on the c.m.s. energy in a collision. This is possible, if one combine only those data that have been measured in more or less the same pseudorapidity intervals. Hence, we suggest to look first of all at ISR~\cite{ISR}, PHENIX~\cite{PHENIX} and ALICE~\cite{ALICE} data that were measured in the most central ($|\eta|<0.8$) pseudorapidity region. 

\begin{figure}[h]
\includegraphics[width =8cm]{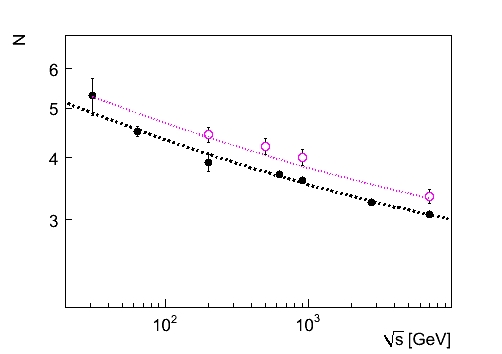}
\caption{\label{fig.2} Variation of the $N$ parameter of (\ref{eq:exppl}) obtained from the fits to the experimental data~\cite{ISR, PHENIX, ALICE} (full points) as a function of c.m.s. energy $\sqrt{s}$ in a collision. Solid line shows a fit of this variation. In addition, open points show the parameters for the data measured in another pseudorapidity interval~\cite{UA11,CMS} and pointed line shows the predictions calculated according to (\ref{eq.n}).}
\end{figure}

Figure~\ref{fig.2} shows the parameter $N$ variation as a function of c.m.s. energy in a collision. One can notice that it can be characterized by the falling $N$-value. It is related to the fact that the probability to produce a high-$p_T$ mini-jet should grow with $\sqrt{s}$. Notably, this behaviour correlates with the fact that $N$ decreases when the rapidity
interval between the secondary hadron and the initial proton increases.

One can extract the following regularity from the fit shown in figure~\ref{fig.2}\footnote{In the~(\ref{eq.sn}) and below $\sqrt{s}$ is given in GeV.}:
\begin{equation}
\label{eq.sn}
N = 2 + 5.25\cdot s^{-0.093}
\end{equation}

Remarkably, in the $s\rightarrow\infty$ limit $N\rightarrow2$, which corresponds to $d^2\sigma/dp^2_T \propto 1/p^4_T$ in the proposed parametrization~(\ref{eq:exppl}). Such behaviour can be expected just from the dimensional counting in pQCD. However, in a real collision one should take the initial conditions and the kinematic restrictions into account, resulting in a higher value of $N$ for lower $\sqrt{s}$. Obviously, such initial conditions should become negligible in the $s\rightarrow\infty$ limit, that is confirmed by the observed behavior~(\ref{eq.sn}).

In addition, figure~\ref{fig.2} shows UA1~\cite{UA11} and CMS~\cite{CMS} data measured under different experimental conditions. In these measurements the pseudorapidity interval was much wider ($|\eta|<2.5$) than in~\cite{ISR,PHENIX,ALICE}. Therefore, one can compare the parameter values obtained from the fit of these data (open points in figure~\ref{fig.2}) to the values extrapolated from (\ref{eq.n}) with $N^0$ calculated according to (\ref{eq.sn}) and $|\eta|=1.25$ (pointed lines) and see a rather good agreement.  Notably, this behaviour correlates with the fact that $N$ decreases when the rapidity interval between the secondary hadron and the same side beam proton increases.

\begin{figure}[h]
\includegraphics[width =8cm]{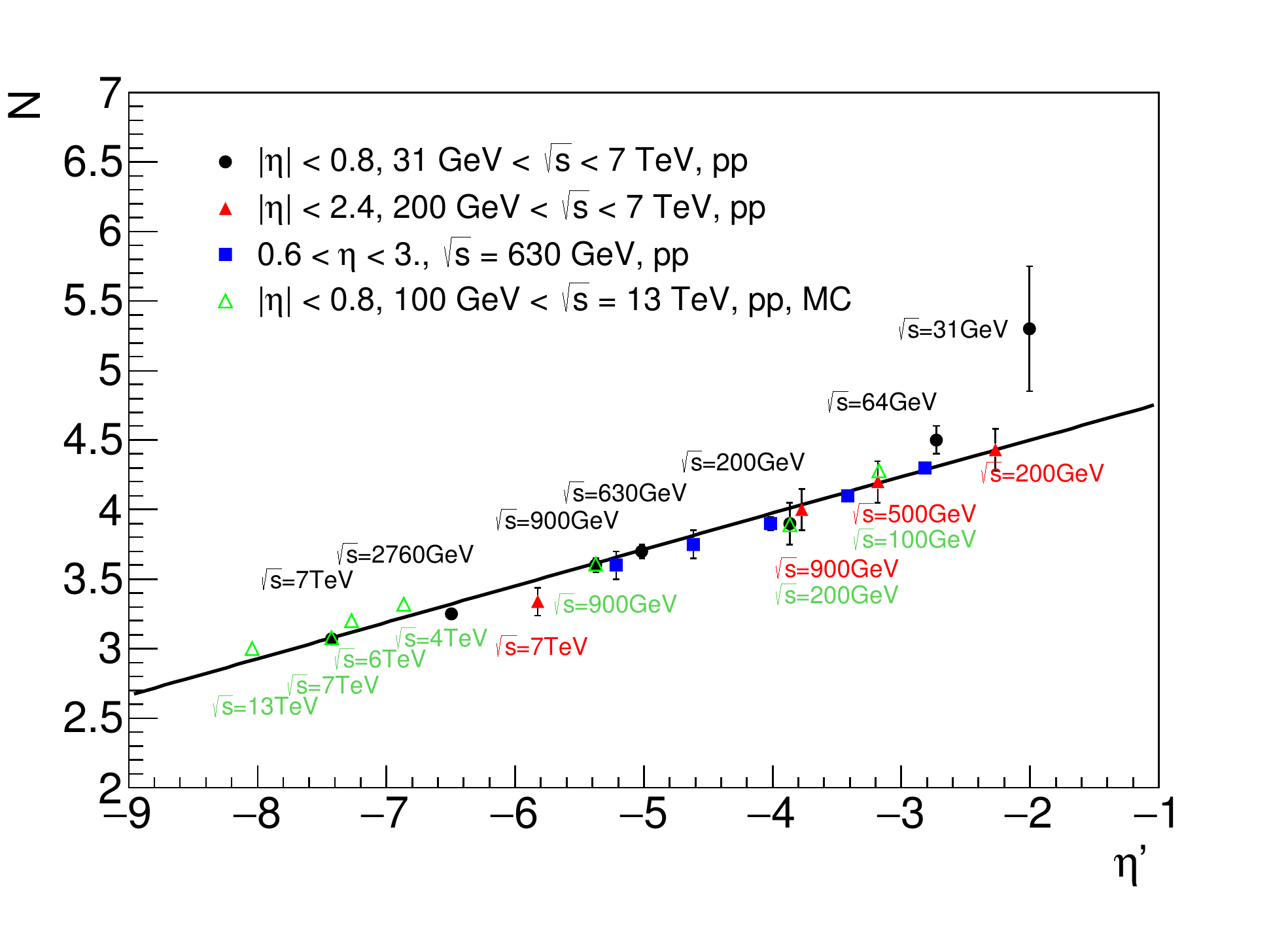}
\caption{\label{fig.N_eta}  The dependence of the N parameter on the pseudorapidity of the secondary hadron in the moving proton rest frame. The data points from different experiments~\cite{UA1, ISR, PHENIX, ALICE, UA11,CMS} are shown together with the generated data(MC).}
\end{figure}

Let us now check this correlation explicitly and calculate the rapidity interval in the moving proton rest frame according to a simple formula:
\begin{equation}
\label{eq.eta'}
\eta' = |\eta| - log(\sqrt{s}/2m_p),
\end{equation}
where $m_p$ is the mass of the incoming proton.
The results of such a procedure are shown in the figure~\ref{fig.N_eta}. Surprisingly, all the points came to a single line in this interpretation. To understand the origin of this universality one might use MC-generators: hard processes at large $p_{T}$ are known to be described by MC generators pretty well, thus it is expected to get the value of N-parameter from the fits of the MC-generated spectra rather close to the real data, but with a higher accuracy and in a wider collision energy range. To check this universality, we have produced the Monte Carlo samples for proton-proton collisions at different energies for inelastic(INEL) events with the PYTHIA 8.2 generator~\cite{pythia}.  Indeed, the values of the parameter N extracted from the fits to the MC-generated spectra are nicely placed at the same line. Thus, a universal parameter describing the shape of the transverse momentum spectra in pp-collisions has been found.

\begin{equation}
\label{eq.eta_eta}
N = 5.04 + 0.27 \eta' 
\end{equation}

\begin{figure}[h]
\includegraphics[width =8cm]{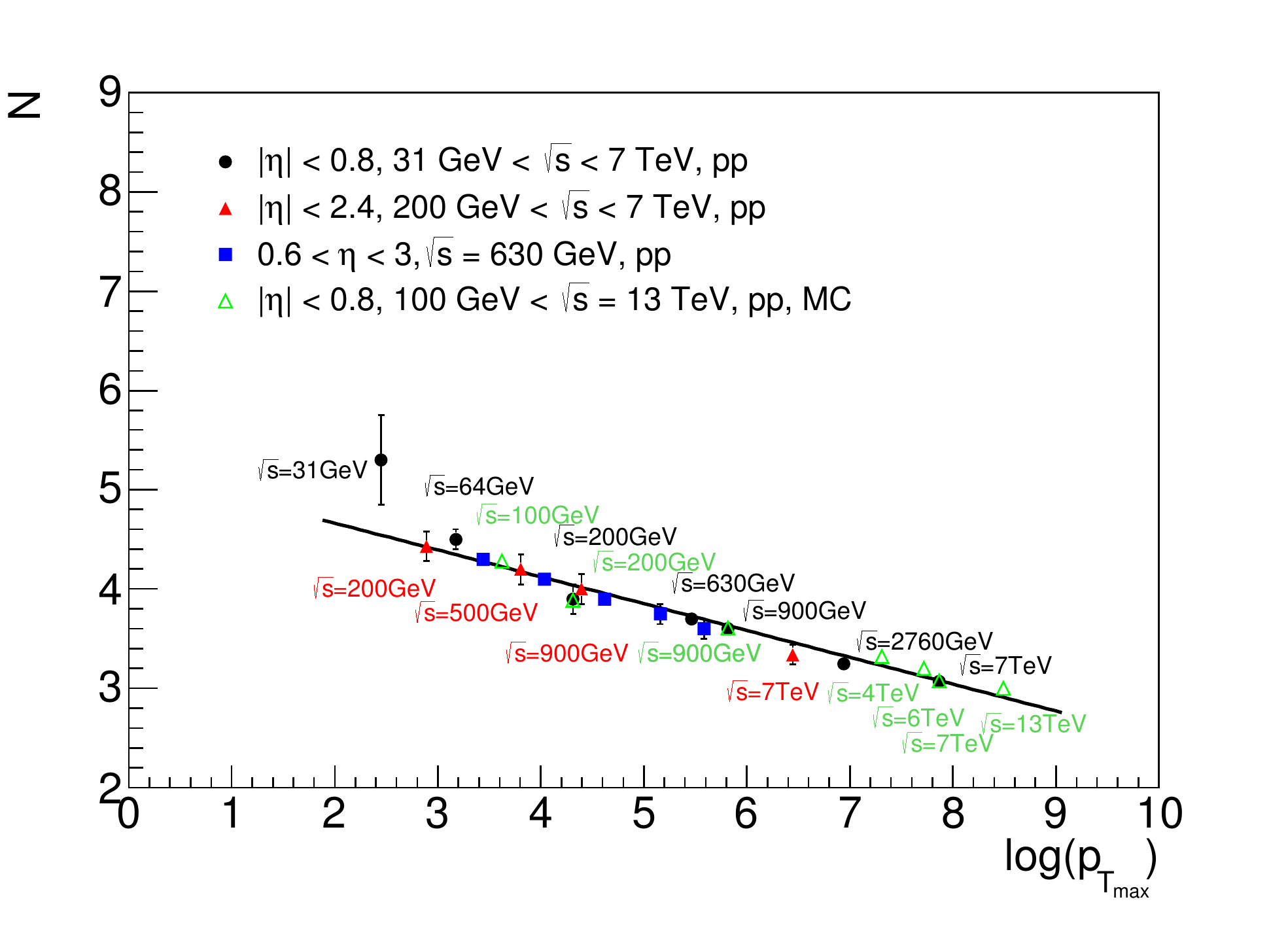}
\caption{\label{fig.N_ptmax}  The dependence of the N parameter on the maximal kinematically allowed transverse momentum of the secondary hadron. The data points from different experiments~\cite{UA1, ISR, PHENIX, ALICE, UA11,CMS} are shown together with the generated data(MC).}
\end{figure}

The further check of the observed phenomena can be made if one plot $N$ as a function of the logarithm of the maximal kinematically allowed transverse momentum $p_{T_{max}}$ of the secondary hadron at the specific pseudorapidity $\eta$.
\begin{equation}
\label{eq.ptmax}
p_{T_{max}} = \frac{\sqrt{s}}{2} sin[2\cdot tan^{-1}(exp(-|\eta|))]
\end{equation}
 This dependence is shown in the figure~\ref{fig.N_ptmax}. One can see, that larger N corresponds to the smaller $p_{T_{max}}$. That should correspond to the $x\rightarrow1$ limit of PDFs, where the fall-off of the perturbative cross section is modified by the fall-off of the PDFs.

\indent Remarkably, similarly to $N$, $T$ and $T_e$ also show dependences as a function of both the collisions energy $\sqrt{s}$ and the measured pseudorapidity interval $\eta$. The variations of the $T$ and $T_e$ parameters were studied in~\cite{Confin}. In~\cite{Confin} the possible theoretical explanation of the thermalized particle production was presented and the following proportionalities were established:
\begin{equation}
\label{eq.t}
T = 409 \cdot (\sqrt{s})^{0.06}\cdot \exp(0.06|\eta|)~MeV
\end{equation}
\begin{equation}
\label{eq.te}
T_e = 98 \cdot (\sqrt{s})^{0.06}\cdot \exp(0.06|\eta|)~MeV
\end{equation}
The parametrizations for $T$ and $T_e$ differes only by a constant factor.
However, both $T$ and $T_e$ parameters reflect the thermalization
which is stronger at higher energies and when closer to the valence
quarks. Therefore, the~(\ref{eq.t},\ref{eq.te}) parametrizations which are functions of centre of mass energy and rapidity interval can be rewritten in the form with only one universal parameter. This universal parameter is the rapidity distance $\eta''$ from the farther incoming proton.\\
\begin{equation}
  \eta''=|\eta| + \log(\sqrt{s}/2m_p)
  \label{eq.eta''}
\end{equation}
\begin{figure}
\includegraphics[width =8cm]{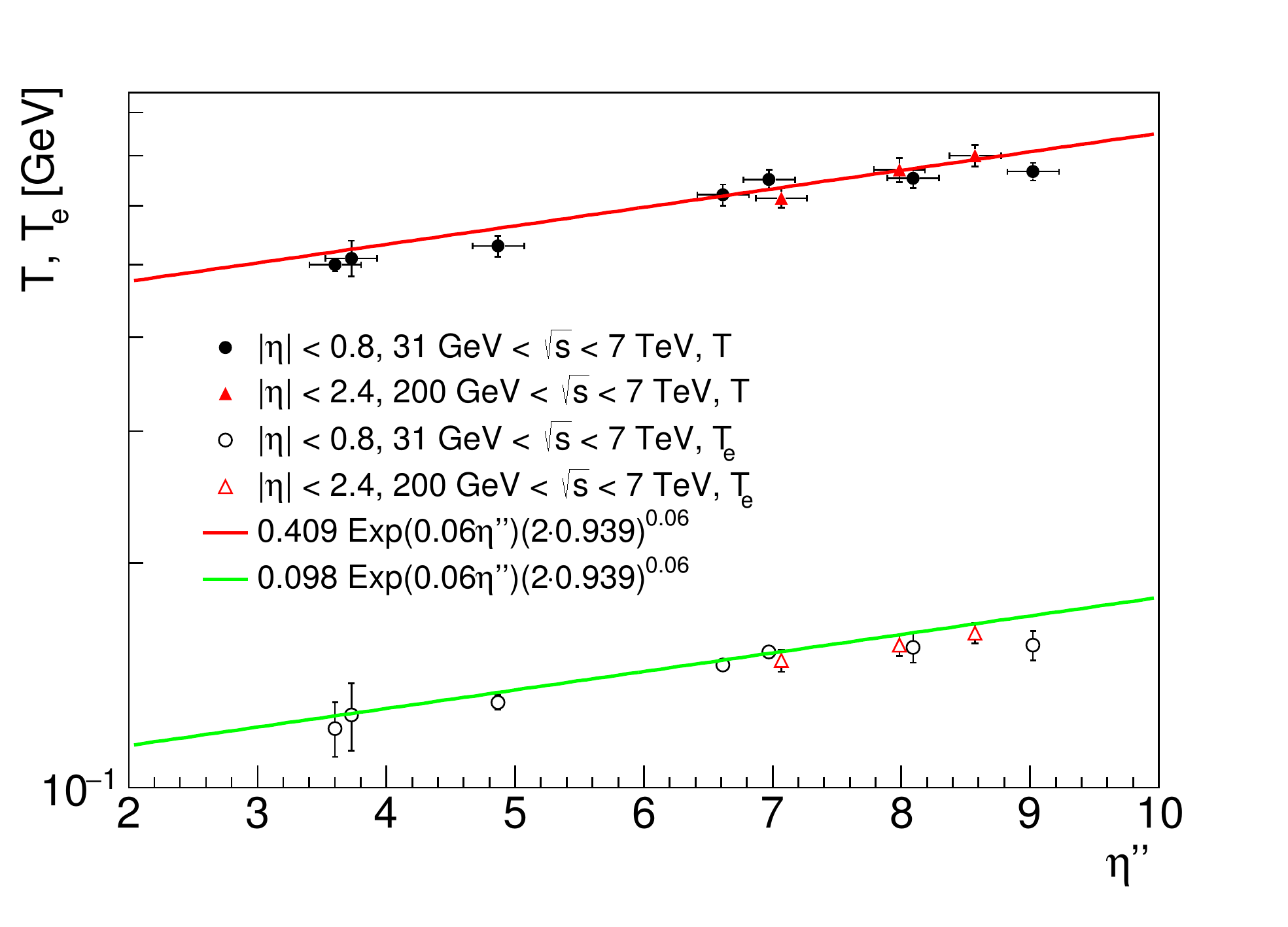}
\caption{\label{fig.TTe}   The dependence of the $T$ and $T_e$ parameter on the
pseudorapidity of the secondary hadron in the moving opposite side proton rest frame. The data points from different experiments~\cite{UA1, ISR, PHENIX, ALICE, UA11,CMS} are shown.}
\end{figure}
Using~(\ref{eq.t},\ref{eq.te}) we get the universal dependence\footnote{In the~(\ref{eq.t_new},\ref{eq.te_new},\ref{eq:final}) $m_p$ is given in units of 1 $GeV/c^2$. }:
\begin{equation}
\label{eq.t_new}
T = 409 \cdot \exp(0.06 \eta'') \cdot (2m_p)^{0.06}~MeV,
\end{equation}
\begin{equation}
\label{eq.te_new}
 T = 98 \cdot \exp(0.06 \eta'') \cdot (2m_p)^{0.06}~MeV.
\end{equation}
The dependences are shown on figure~\ref{fig.TTe}.

\section{Prediction for further measurements }
Though MC was shown to provide a nice description of the high-pt part of the spectra,  the nature of the soft particle production still remains ambiguous and varies for different MC-generators. Therefore, the next step in understanding the underlying dynamics of high energy hadronic processes was done in a recent analysis~\cite{Confin}.

In~\cite{OURR} it was shown, that the introduced approach is able to give predictions on the pseudorapidity distributions in high energy collisions for non-single diffractive events (NSD). Using the parameterizations from~\cite{OURR} in addition with (\ref{eq.n})-(\ref{eq.te}) one can provide a formula that describes the shapes of charged particles spectra, being function only of a centre of mass energy and a measured pseudorapidity region. Let us now summarize all the equations to obtain the final result\footnote{In the~(\ref{eq.gaus}) and below $A$ and $A_e$ is measured in mb/$GeV^{2}$, $A_{pow}$ and $A_{exp}$ in mb}.\\

\begin{equation}
\left(\frac{d\sigma}{d\eta} \right)_{power} = A_{power} \cdot \exp \left(- \frac{\eta^2}{2 \sigma_{power}^2} \right),
\label{eq.gaus}
\end{equation}

\begin{equation}
\begin{split}
\left(  \frac{d\sigma}{d\eta} \right)_{exp} = A_{exp} \cdot \exp \left(-\frac{(\eta - \eta_{exp})^2}{ 2 \sigma_{exp}^2} \right)  + \\ A_{exp} \cdot \exp \left (- \frac{(\eta + \eta_{exp})^2}{2 \sigma_{exp}^2} \right),
\end{split}
\label{eq.gaus2}
\end{equation}

\begin{equation}
\label{eq.sigpl}
\sigma_{power} = 0.217 + 0.235\cdot \ln\sqrt{s},
\end{equation}
\begin{equation}
\label{eq.eta}
\eta_{exp} = 0.692 + 0.293 \cdot \ln \sqrt{s},
\end{equation}
\begin{equation}
\label{eq.sigexp}
\sigma_{exp} = 0.896 + 0.136\cdot \ln \sqrt{s},
\end{equation}
\begin{equation}
\label{eq.apl}
A_{power} = 0.13\cdot s^{0.175},
\end{equation}
\begin{equation}
\label{eq.aexp}
A_{exp} = 0.76\cdot s^{0.106},
\end{equation}
Now, with the knowledge of the variations of $T$, $N$ and $T_e$ parameters and the exponential and power-law contributions one can calculate the normalisation parameters $A$ and $A_e$ in (\ref{eq:exppl}) in the following way.\\

\begin{equation}
\frac{d\sigma}{d\eta} = \left( \frac{d\sigma}{d\eta} \right)_{exp} + \left( \frac{d\sigma}{d\eta} \right)_{pow} = \int_{0}^{\infty} \frac{d^2\sigma}{d\eta dP_T^2} dp_T^2
\end{equation}	

 \begin{equation}
\label{eq.si2}
\left( \frac{d\sigma}{d\eta} \right)_{pow} = \int_{0}^{\infty}\frac{A}{(1+\frac{p_T^2}{T^{2}\cdot N})^N} dp_T^2 = \frac{A N T^2}{N - 1}
\end{equation}
 
 \begin{equation}
\label{eq.si3}
\left(\frac{d\sigma}{d\eta}\right)_{exp} = \int_{0}^{\infty} A_e\exp {(-E_{Tkin}/T_e)} dp_T^2 = 2 A_e T_e (m + T_e)
\end{equation}

\begin{figure}[h]
\includegraphics[width =8cm]{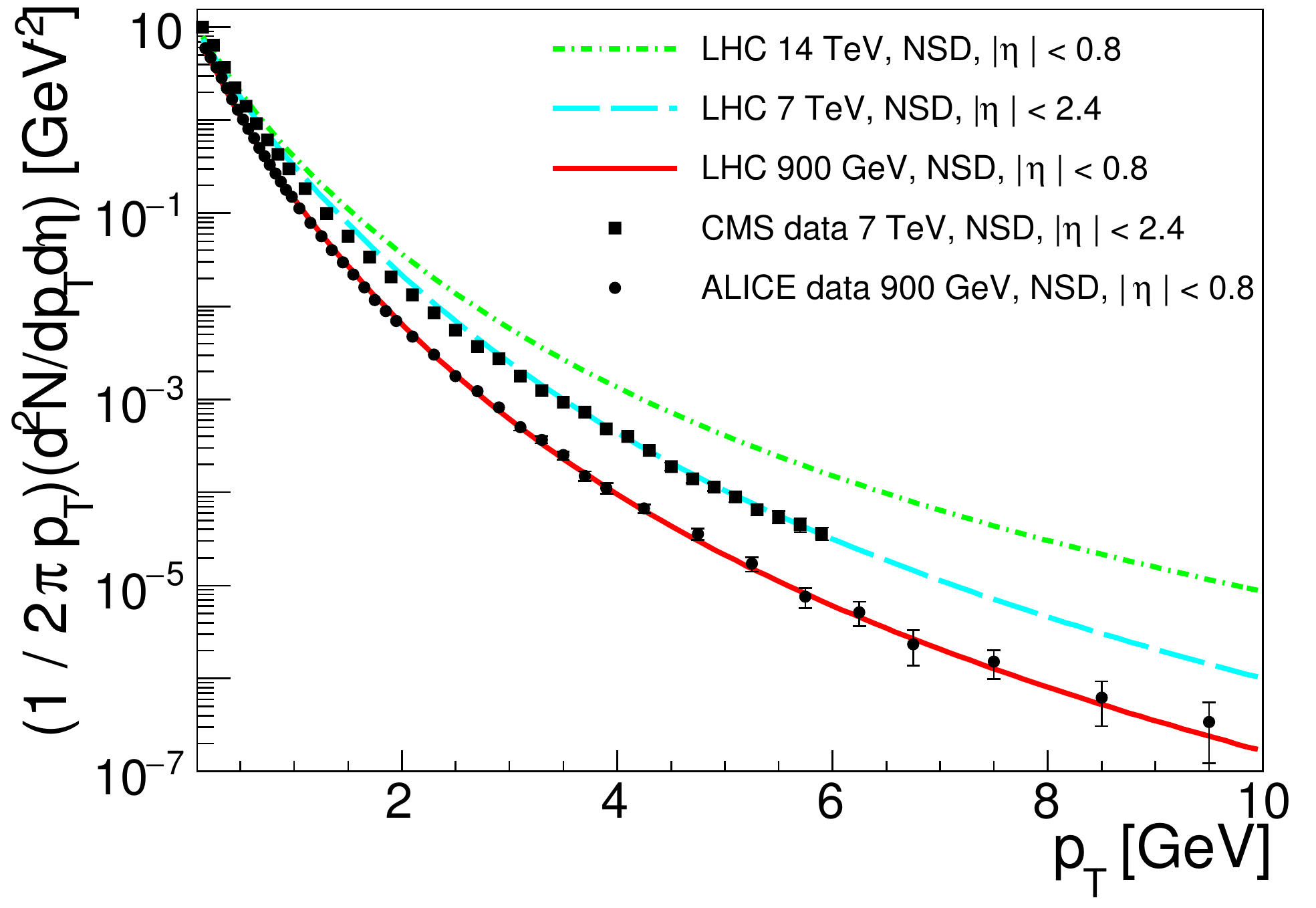}
\caption{\label{fig.3}Predictions the yield of charged particles $(1/2 \pi p_{T})d^2N/(d\eta dp_T)$ in high energy collisions in NSD events together with data points from ALICE~\cite{Alice900} and CMS~\cite{CMS} experiments.}
\end{figure}

Thus, we get the set of equations allowing us to make a prediction for a double differential cross section, using the formula (\ref{eq:exppl}):

\begin{equation}
\label{eq:final}
 \begin{cases}
|\eta| = (|\eta|_{max} + |\eta|_{min})/2,\\
N = 5.04 + 0.27 \eta' \\
T = 409 \cdot \exp(0.06 \eta'') \cdot (2m_p)^{0.06} ,\\
 T = 98 \cdot \exp(0.06 \eta'') \cdot (2m_p)^{0.06} ,\\
A_e = \frac{1}{2T_e (m+T_e)}(d\sigma / d\eta)_{exp},\\
A = \frac{(N - 1)}{N T^2}(d\sigma / d\eta)_{pow},
\end{cases}
\end{equation}

where $\eta'$ and $\eta''$can be calculated using~(\ref{eq.eta'}) and ~(\ref{eq.eta''}) correspondingly.\\
\indent Now, one can calculate double differential cross sections $d^2\sigma/(d\eta dp_T^2)$ of charged particle production in high energy collisions at different energies for NSD events. These predictions are shown in figure~\ref{fig.3} for $|\eta|< 0.8$  and $|\eta|< 2.4$  pseudorapidity intervals together with the experimental data measured by ALICE~\cite{Alice900} and CMS~\cite{CMS}. A good agreement of the prediction with the data can be observed. Thus, these results~(\ref{eq:final}) give us a powerful tool for predicting the spectral shapes in NSD events. 

\section{Conclusion}
In conclusion, transverse momentum spectra in $pp$-collisions have been considered using a two component model. Variations of the parameters obtained from the fit have been studied as a function of pseudorapidity $\eta$ and c.m.s. energy $\sqrt{s}$ in the collision. A universal parameter describing a shape of the spectra in pp-collisions was found to be a preudorapidity of a secondary hadron in moving proton rest frame. Finally, the observed dependences, together with previous investigations allowed to make predictions on double differential spectra $d^2\sigma /(dp_T^2d\eta)$ at higher energies, successfully tested on the available experimental data. \\

\begin{acknowledgements}
The authors thank Prof. Mikhail Ryskin, Prof. Torbjorn Sjostrand and Dr. Peter Skands for fruitful discussions and help provided during the preparation of this paper. 
\end{acknowledgements}


\begin{thebibliography}{99}
\bibitem{OUR1}
  A.~A.~Bylinkin and A.~A.~Rostovtsev,
  Phys.\ Atom.\ Nucl.\  {\bf 75} (2012) 999
   Yad.\ Fiz.\  {\bf 75} (2012) 1060;\\
A.~A.~Bylinkin and A.~A.~Rostovtsev,
  arXiv:1008.0332 [hep-ph].

\bibitem{UA1}
G.~Bocquet {\it et al.} [UA1 Collaboration],
Phys.Lett.B{\bf 366}:434-440,1996


\bibitem{OURM}
  A.~A.~Bylinkin, M.~G.~Ryskin,
  arXiv:1404.4739 [hep-ph].

\bibitem{OURR}
  A.~A.~Bylinkin {\it et al.}  [Institute for Theoretical and Experimental Physics, ITEP, Moscow, Russia Collaboration],
  arXiv:1404.7302 [hep-ph].
  
\bibitem{Confin}
Alexander A. Bylinkin, Dmitri E. Kharzeev, Andrei A. Rostovtsev,
 arXiv:1407.4087 [hep-ph]

\bibitem{ISR}
  K.~Alpgard {\it et al.}  [UA5 Collaboration],
  Phys.\ Lett.\ B {\bf 112} (1982) 183.

\bibitem{pythia}
T. Sjostrand, S. Ask, J. R. Christiansen, R. Corke, N. Desai, P. Ilten, S. Mrenna, S. Prestel, C. O. Rasmussen, P. Z. Skands,
arXiv:1410.3012  [hep-ph]


\bibitem{PHENIX}
  A.~Adare {\it et al.}  [PHENIX Collaboration],
  Phys.\ Rev.\ C {\bf 83} (2011) 064903
  [arXiv:1102.0753 [nucl-ex]].

 \bibitem{Alice900}
K.~ Aamodt {\it et al.}  [ALICE Collaboration],
Phys.\ Lett.\ B {\bf 693} (2010) 53-68
 [arXiv:1007.0719 [hep-ex]].


\bibitem{ALICE}
  B.~B.~Abelev {\it et al.}  [ALICE Collaboration],
  Eur.\ Phys.\ J.\ C {\bf 73} (2013) 2662
  [arXiv:1307.1093 [nucl-ex]].
 
\bibitem{UA11}
  C.~Albajar {\it et al.}  [UA1 Collaboration],
  Nucl.\ Phys.\ B {\bf 335} (1990) 261.

\bibitem{CMS}
  V.~Khachatryan {\it et al.}  [CMS Collaboration],
  Phys.\ Rev.\ Lett.\  {\bf 105} (2010) 022002
  [arXiv:1005.3299 [hep-ex]].

\end{thebibliography}
\end{document}